\documentclass[aps, prl,superscriptaddress,showpacs,twocolumn,10pt]{revtex4-1}
\pdfoutput=1

\usepackage{amsfonts,amssymb,amsmath}
\usepackage{graphics,graphicx,epsfig}
\usepackage{amsthm}
\usepackage{placeins}					
\usepackage{enumitem} 

\renewcommand{\eqref}[1]{Eq.~(\ref{#1})}
\newcommand{\fref}[1]{Fig.~\ref{#1}}
\newcommand{\tref}[1]{Tab.~\ref{#1}}

\begin{document}

\title{Nanometer scale resolution, multi-channel separation of spherical particles in a rocking ratchet with increasing barrier heights}

\author{Philippe Nicollier}
\thanks{These authors contributed equally to this work.}
\affiliation{IBM Research -- Zurich, S\"aumerstrasse 4, 8803 R\"uschlikon, Switzerland}

\author{Christian~Schwemmer}
\thanks{These authors contributed equally to this work.}
\affiliation{IBM Research -- Zurich, S\"aumerstrasse 4, 8803 R\"uschlikon, Switzerland}

\author{Francesca Ruggeri}
\affiliation{IBM Research -- Zurich, S\"aumerstrasse 4, 8803 R\"uschlikon, Switzerland}

\author{Daniel~Widmer}
\affiliation{IBM Research -- Zurich, S\"aumerstrasse 4, 8803 R\"uschlikon, Switzerland}

\author{Xiaoyu Ma}
\affiliation{IBM Research -- Zurich, S\"aumerstrasse 4, 8803 R\"uschlikon, Switzerland}

\author{Armin~W.~Knoll}
\email{ark$@$zurich.ibm.com}
\affiliation{IBM Research -- Zurich, S\"aumerstrasse 4, 8803 R\"uschlikon, Switzerland}

\begin{abstract}
	We present a nanoparticle size-separation device based on a nanofluidic rocking Brownian motor. It features a ratchet-shaped electrostatic particle potential with increasing barrier heights along the particle transport direction. The sharp drop of the particle current with barrier height is exploited to separate a particle suspension into multiple sub-populations. By solving the Fokker--Planck equation, we show that the physics of the separation mechanism is governed by the energy landscape under forward tilt of the ratchet. For a given device geometry and sorting duration, the applied force is thus the only tunable parameter to increase the separation resolution. For the experimental conditions of $3.5$\,V applied voltage and 20\,s sorting, we predict a separation resolution of $\sim 2$\,nm, supported by experimental data for separating spherical gold particles of nominal 80 and 100\,nm diameters.
\end{abstract}


\date{\today}
\maketitle

Separation of nanoparticles and molecules is a highly relevant technical task~\cite{salafi2017advancements}, for which the resolution is typically limited by diffusion. Therefore, high driving fields are required to enhance the resolution. For example, in capillary electrophoresis~\cite{swerdlow1990capillary} or sieving devices based on nanoscale gaps~\cite{fu2007patterned}, electric DC fields of several $100\,$V/cm are typically used, thus limiting applications in mobile or lab-on-chip devices. Similarly, high pressures are required in deterministic lateral displacement arrays~\cite{Huang14052004, wunsch2016nanoscale}. 

Brownian motor-based devices were envisioned for particle transport and separation~\cite{Haenggi2009} as early as the 90s. In contrast to the methods mentioned above, Brownian motors transport particles with an AC modulation of either an asymmetric potential~\cite{Rosselet1994Nature, Gorre-Talini1997} (flashing ratchets) or a driving force combined with a static ratchet potential~\cite{Magnasco1993, Bartussek1994, faucheux1995selection, reimann2002introduction} (rocking ratchets). For flashing ratchets, the linearly decreasing diffusion coefficient with particle size provides a separation mechanism~\cite{faucheux1995selection, Gorre-Talini1997}. Rocking ratchets exhibit a highly non-linear particle current as function ~\cite{Bartussek1994} of the applied force and frequency, which was suggested to be useful for particle separation~\cite{reimann2002introduction, Bartussek1994}. Recently, we implemented a rocking Brownian motor for nanoparticles~\cite{skaug2018nanofluidic, Schwemmer2018Experimental}. We demonstrated the separation of gold spheres measuring 60 and 100\,nm in diameter within a few seconds. The device footprint was small, i.e.~less than $20\,\mu$m, enabling high fields with voltages of less than 5\,V and stable operation over hours. The separation mechanism was based on two intercalated Brownian motors pointing in opposite directions. Particles of different size preferably occupied one of the two motors and were therefore extracted to opposite ends of the device. Modeling suggested that the same separation device is capable of separating two gold sphere populations with $\approx 1\,$nm difference in radius~\cite{skaug2018nanofluidic}.

Here, we present a separation device based on a rocking ratchet that splits a particle population into several sub-populations with similar resolution. The device separates the particles by transporting them across increasing potential barriers in the ratchet direction. The separation mechanism is thus markedly different from our previous implementation. It exploits the ``Arrhenius-like'' onset of the particle current with decreasing ratchet potential barriers, which was simulated by Bartussek et al. and suggested as a separation mechanism for particles with similar diffusion coefficients~\cite{reimann2002introduction, Bartussek1994}.

In the following, we first describe the experiment and observe the particle current in the device for gold particles nominally 60\,nm in diameter. The results agree well with a numerical solution of the Fokker--Planck equation using measured physical parameters as input. Simulations allow us to assess the resolution of the sorting device and its scaling with separation time and force. For the experimental parameters used, we obtain a resolution of 2 nm (4\,$\sigma$), which we compare with the experimental resolution based on microscopic inspection after particle deposition. 

\begin{figure}[h]
	\centering
	\includegraphics[width=0.48\textwidth]{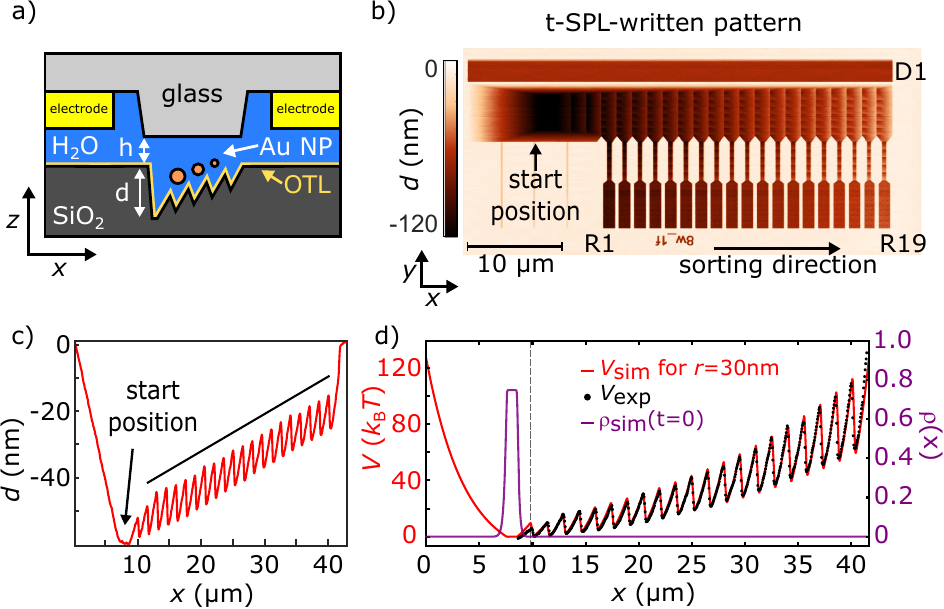}
	\caption{(color online) (a)~Schematic side view of the nanofluidic slit (not to scale). A central pillar of $\approx 40\,\mu$m height was etched from a cover slip and surrounded by four Cr--Au electrodes.
		The ratchet structure in SiO$_2$ was covered with a thin layer of OTL polymer. 
		(b)~Topography of the sorting device in PPA after t-SPL patterning. 
		From the inclined ratchet (top), particles can be driven into reservoirs (R1 to R19) by linear ratchets after being sorted. Force and diffusivity are measured in D1.  	
		(c)~Cross section through the device after transfer to SiO$_2$. The 19 teeth in the sorting ratchet span a vertical distance of 28\,nm, corresponding to $\Delta z = 1.6\,$nm per tooth. (d)~Extrapolated potential $V_{exp}(x)$ for a gap distance of $64 \pm 2\,$nm (black dots) and the input potential $V_{sim}(x)$ for simulating the device (red line). We used an initial probability density $\rho_0(x) \propto \exp(-V(x)/k_{\rm B}T)$ with $0 < x < 10\mu$m (dashed line) for our simulations (violet).}
	\label{fig:fig1}
\end{figure}

\emph{Experimental implementation.}---The potential landscape experienced by the particles in our nanofluidic device is based on the electrostatic interaction of charged particles with like-charged walls~\cite{Krishnan10nature}, see \fref{fig:fig1}a). The schematically depicted ratchet with increasing tooth height leads to the aforementioned increase in energy barriers. We note that a similar geometry was used recently for nanofluidic size exclusion~\cite{liao2018subnanometer}. We used $60$\,nm gold spheres (EM.GC60 from BBI solutions, batch $\# 19080123$) to study the particle transport in the device. A volume of $\approx30\,\mu$l of the suspension was deposited on the sample and subsequently confined to a tunable nanofluidic slit using the nanofluidic confinement apparatus (NCA), see SM1 and SM2 of the Supplemental Material~(SM)~\cite{SM} and Refs.~\cite{fringes2018nanofluidic,skaug2018nanofluidic} for details. 
The top boundary of the slit consisted of a cover glass with a central mesa of height $\approx 40\,\mu$m. The lower boundary was a silicon chip~\cite{SM} with thermal oxide thickness of $\approx235$\,nm, see \fref{fig:fig1}a). The geometry of the device was patterned by thermal scanning-probe lithography (t-SPL)~\cite{Garcia2014Advanced} into polyphthalaldehyde (PPA), see \fref{fig:fig1}b), and then dry-etched into the $\rm SiO_2$ layer~\cite{rawlings2017control}. In the final device, the average height difference between each of the 19 neighboring teeth was $1.6$\,nm, see \fref{fig:fig1}c).
Finally, the sample was coated with a $10$-nm-thick organic transfer layer (OTL, PiBond Oy) of a polymeric material required to immobilize the particles on the sample surface after sorting~\cite{fringes2019Deterministic}. A more detailed description of the sample preparation can be found in SM3~\cite{SM}. For imaging, we used interferometric scattering detection (iSCAT)~\cite{Lindfors04prl, fringes2016situ}, recording at 250 frames per second. A temporal stability of the nanofluidic gap of $\approx 2\,$nm RMS (see SM4~\cite{SM}) was measured using this detection scheme, limited by the stability of our laser source.

%
%
%
%
%
%
The interaction potential $W(x,y,z)$ of charged nanoparticles of radius~$r$ in a gap of height~$h$ and pattern depth~$d(x,y)$ can be approximated by the sphere--plane interaction~\cite{fringes2016situ, skaug2018nanofluidic, Behrens2001} 
\begin{equation}
W_0 r \psi_{S} \left(\psi_{P,1} e^{-\kappa(z-r)}+\psi_{P,2} e^{-\kappa(h+d(x,y)-z-r)} \right),
\label{eq:potential}
\end{equation}
where $W_0=4 \pi \epsilon \epsilon_0$, $\epsilon$ and $\epsilon_0$ are the relative and the vacuum permittivities,  $\psi_{S}, \psi_{P,1}, \psi_{P,2}$ are the effective surface potentials of the sphere and the two planes, $z$ is the vertical particle position measured from the substrate interface, and $\kappa^{-1}$ is the Debye length, see \fref{fig:fig1}a) and~\cite{skaug2018nanofluidic}. Note that \eqref{eq:potential} only holds in the case of a topography with shallow slopes (i.e.~$<\,1$) as in our sample. The 2D occupation probability $P(x,y)$ is  obtained by integrating the 3D probability density $p(x,y,z) \propto e^{- W(x,y,z)/(k_{\rm{B}} T)}$
\begin{equation}
P(x,y) = \left( C \int_{z=r}^{z = d(x,y) + h - r} e^{- W(x,y,z)/(k_{\rm{B}} T) } dz \right),
\label{eq:prop2D}
\end{equation}
where $k_{\rm{B}} T$ is the thermal energy and $C$ is a normalization constant. The free energy $V(x,y)$ up to a reference potential $V_0$ including positional entropy is given by 
\begin{equation}
V(x,y) = -\ln (P(x,y)) + V_0.
\label{eq:free_energy}
\end{equation}

In order to quantify $V_{\rm exp}(x)$  experimentally along the ratchet direction $x$, we trapped $\approx 30$ particles in the ratchet area during the approach of the glass pillar. Their $x$-coordinates were tracked for three gap distances of $h = 151 \pm 2$\,nm, $h = 112 \pm 2$\,nm, and $h = 103 \pm 2$\,nm using Trackpy for Python~\cite{SM, Trackpy, crocker1996methods}. Identifying the normalized frequency of  particles observed at position $x$ with the particle occupancy probability $P_{\rm exp} (x)$ and assuming $\psi_{P,1} = \psi_{P,2} = \psi_{P}$, we inferred from a global fit a Debye length of $\kappa^{-1} = 10.8 \pm 0.1$\,nm and $2 W_0 \psi_{S} \psi_{P} = 5.3 \pm 0.2\,k_{\rm B}T/{\rm nm}$. Using~\eqref{eq:free_energy}, we then extrapolated $V_{\rm exp}(x)$ to other gap distances, see also SM5~\cite{SM}. 
\newline
\indent
Size separation experiments were performed at a gap distance of $h=64 \pm 2$\,nm, for which $V_{exp}(x)$ is shown in \fref{fig:fig1}d).   
\begin{figure}[ht]
	\centering
	\includegraphics[width=0.48\textwidth]{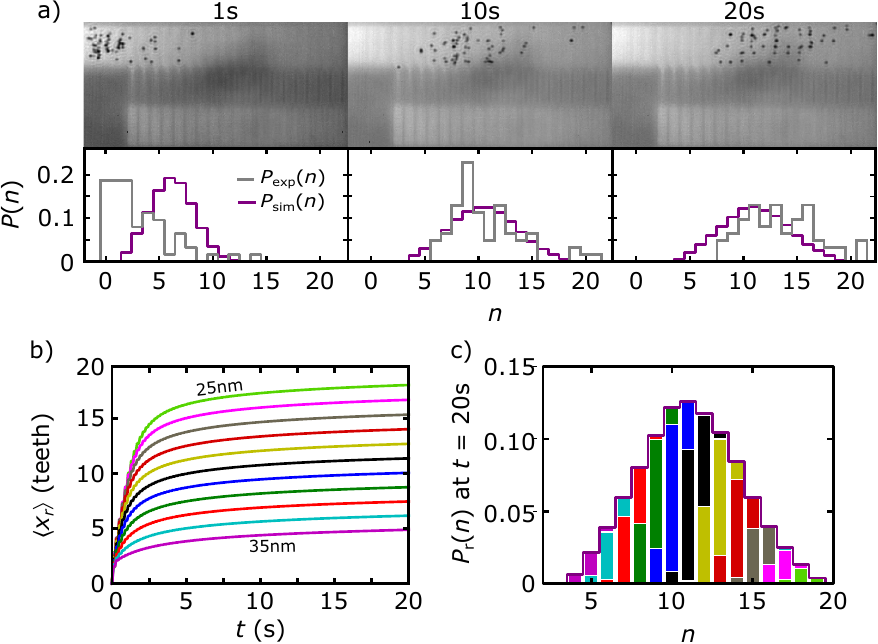}
	\caption{(color online) (a)~Top: iSCAT images of the sorting process after 1, 10, and 20\,s of an applied AC voltage of $3.5\,$V at $5$\,Hz. Bottom: Comparison of the experimental, $P_{\rm exp}(n)$, and simulated, $P_{\rm sim}(n)$, particle distributions. (b)~Expected average travel distance for particles of radii between $25$ and $35\,$nm with $1\,$nm difference in radius as a function of time. (c)~Simulated probability distribution $P_{\rm r}(n)$ after 20\,s of sorting. The width of the distributions of single particle types, marked by separate colors, indicates the simulated resolution of the sorting device. The color code corresponds to the one used in~(b).}
	\label{fig:fig2}
\end{figure}
The particles were first transported to the starting position of the ratchet, see \fref{fig:fig1}c), which corresponds to the lowest energy in the system. In order to speed up the process we exploited the previously observed reversal of the particle current in rocking Brownian motors~\cite{Schwemmer2018Experimental} at rocking frequencies above 150\,Hz. Using $2$ to $3$\,V and a frequency of $300$ to $500$\,Hz across each pair of electrodes, the particles were transported to the starting position within a few minutes.

Next, we applied an AC voltage of $3.5$\,V amplitude at a frequency of $5$\,Hz in $x$ direction. A second voltage of $3$\,V amplitude at 500\,Hz along the $y$-axis prevented particles from diffusing into the reservoirs, again exploiting current reversal~\cite{Schwemmer2018Experimental}. Three representative frames of the transport process are shown in \fref{fig:fig2}a) together with the measured particle occupancy for each tooth in the ratchet. For particles entering the ratchet, the observed particle speed was high, and slowed down dramatically after $\approx 5$\,s. From 10 to 20\,s, the population shifted on average by just one tooth. In this state, the particle population was spread over 13 teeth, indicating a fine separation of particles.
\newline
\indent
\emph{Modelling.}---
The dynamics of a Brownian motor with ratchet potential $V(x)$
and a external rocking force $F(t)$ can be expressed in terms of a probability density $\rho(x,t)$, 
which obeys the Fokker--Planck equation~\cite{Risken1989} 
\begin{equation}
\partial_t \rho(x,t) = \partial_x \left[ \left( \frac{1}{\gamma}\partial_x \tilde{V}(x,t) + D_0 \partial_x \right) \rho(x,t) \right],
\label{eq:fokkerplanck}
\end{equation}
where $\gamma$ is the drag constant,  $D_0$ the diffusion coefficient, and  $\tilde{V}(x,t) = V(x) + x F(t) $ tilted potential. 
The sorting process represents an initial value problem where $\rho_0(x)$ 
of the unsorted particles at $t=0$ is transformed into $\rho_{\rm final}(x)$ of the sorted particles. The propagation of $\rho_0(x) \longrightarrow \rho_{\rm final}(x)$ is given by \eqref{eq:fokkerplanck} and can normally be calculated  only numerically. Therefore, we discretized \eqref{eq:fokkerplanck} with respect to $x$ and approximated the spatial derivatives by finite differences.
The resulting system of ordinary differential equations was then computed by standard solvers for 
ordinary differential equations. For more details on the numerical solution of \eqref{eq:fokkerplanck}, see SM6~\cite{SM}.
We used the extrapolated interaction potential as input, see \fref{fig:fig1}d), and a rocking force $F$ that was inferred~\cite{skaug2018nanofluidic} from the average drift speed $\langle v_{\rm drift} \rangle$ of the particles in the 30-nm-deep drift field D1 (see \fref{fig:fig1}b) of $F = k_{\rm B}T \langle v_{\rm drift} \rangle/D_0 = 20.7 \pm 1.3\,k_{\rm{B}} T / \mu {\rm m}$. The average diffusion constant ${D_0 = 3.2 \pm 0.2\,\mu {\rm m}^2/{\rm s}}$ was determined for particles in field D1 without applied fields. 
\indent

According to the manufacturer, the mean radius of the gold nanoparticles is $30.3\,$nm, and their coefficient of variation is $8\%$. Assuming a Gaussian size distribution, this results in $97.5\%$ of the particles having a radius of between $25$ and $35\,$nm. We simulated particle sizes in this range with a radial difference of 1\,nm, and scaled $V(x)$ (\eqref{eq:potential}), $F \propto r$ and $D_0 \propto r^{-1}$, accordingly\cite{skaug2018nanofluidic}. Each particle species was simulated separately. The resulting probability densities were then summed up and weighted with their corresponding relative portion in the original dispersion, see also SM7~\cite{SM}. The resulting particle probability distribution $P_{\rm sim}(x)$ after 1, 10 and 20\,s of sorting can be seen in \fref{fig:fig2}a). This agrees well with the experimentally observed evolution of $P_{\rm exp}(n)$.

The temporal evolution of the mean travel distance $\left\langle x_r \right\rangle$ for particles with different radii $r$ is shown in \fref{fig:fig2}b) and reflects the observation of Bartussek et al.~\cite{Bartussek1994} of an Arrhenius-like decrease of the particle current with increasing energy barriers. After a fast transport into the ratchet, the average speed decreases sharply, and the particles enter a quasi-steady-state. For particles of different sizes, the transition occurs at a different tooth number because smaller particles experience less interaction potential for the same tooth height, see~\eqref{eq:potential}. As a result, particles of different sizes are transported to different locations in the device, see~\fref{fig:fig2}c). 


\begin{figure}[ht]
	\centering
	\includegraphics[width=0.48\textwidth]{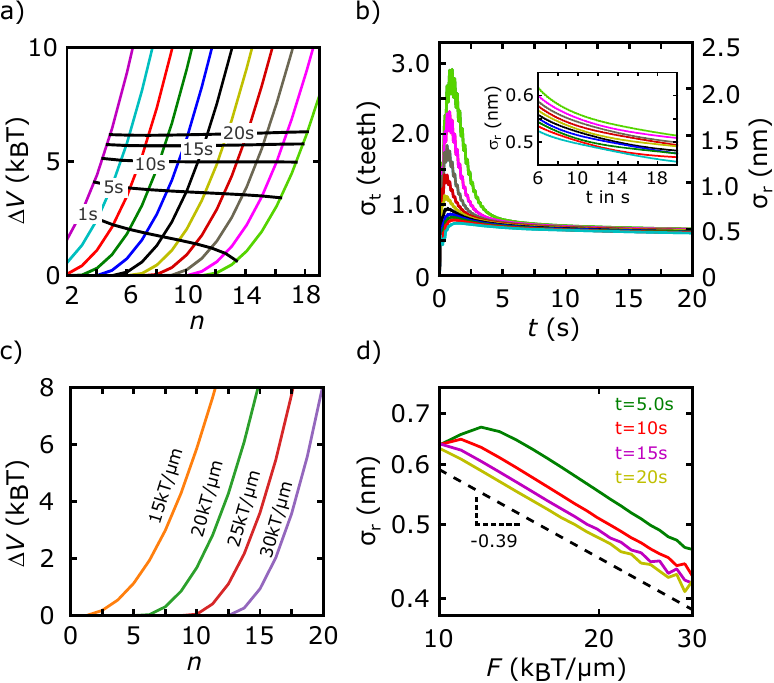}
	\caption{(color online) (a)~Forward bias potential barriers for particles with radius 35 to 25~nm (left to right) and a force of $20.7\,k_{\rm{B}} T / \mu$m (colored lines). The intercept with the black lines marks the average travel distances and reached barrier heights after 5, 10, 15 and 20\,s. (b)~Standard deviation $\sigma_r$ of the radius distribution in the central tooth for each particle size. The inset shows a close-up for longer sorting durations. (c)~Remaining energy barriers under forward bias for  particles with a radius of 30~nm and applied forces of 15, 20, 25, and 30$\,k_{\rm{B}} T/\mu$m. (d)~$\sigma_r$ as a function of applied force and different separation durations.}
	\label{fig:fig3}
\end{figure}

The behavior of the particle transport into the device is almost entirely controlled by the energy landscape experienced by the particles in the forward bias case. The backward current is significant only in the first 2.5~s of the experiment, and then becomes negligible for all considered particle sizes, see SM8~\cite{SM}. 
The forward energy barriers are plotted in \fref{fig:fig3}a). Depending on the particle size, the barriers start to deviate from zero at different tooth numbers. After sorting durations of 5 to 20~s, the particles on average arrive at energy barriers between 3 and 6~$k_{\rm{B}} T$. 

The rapidly increasing energy barriers lead to a focusing of the particle density. This can be seen from the standard deviation $\sigma_t$ of the particle distribution across the ratchet teeth measured for each particle population as shown in \fref{fig:fig3}b). $\sigma_t$ increases within the first few seconds and then decreases to less than one tooth after $t \approx 5\,$s. $\sigma_t$ is related to the spread $\sigma_r$ of particle radii in a given tooth (different colors per tooth in \fref{fig:fig2}c)), and we find $\sigma_r\,/$nm$ = 0.76\,\sigma_t$ shown as the right-hand $y$-axis in \fref{fig:fig3}b), for details see SM9~\cite{SM}. $\sigma_r$  rapidly approaches a value of less than 0.6\,nm after 5~s, and then decreases slowly to $\sim 0.5$\,nm after 20~s. 
If we define the resolution~\cite{derenyi1998ac} of the device to be $4\,\sigma_r$, it follows that we can separate two particles with a difference in radius of 2\,nm.

As $\sigma_r$ depends exclusively on the forward-biased energy landscape, the only tunable parameter to increase the resolution in a device, for a given time span, is the amplitude of the force. A higher force leads to more steeply increasing energy barriers as a consequence of their exponential scaling with $x$, see \fref{fig:fig3}c). 
As a result, the separation resolution increases. This can be seen from the decrease of $\sigma_r$ shown in \fref{fig:fig3}d). For this device, the sorting resolution scales roughly as a power-law  $F^{-0.39}$ and reaches $\sigma_r \approx 0.4$\,nm at 30 $k_{\rm{B}} T / \mu m$.

\begin{figure}[ht]
	\centering
	\includegraphics[width=0.48\textwidth]{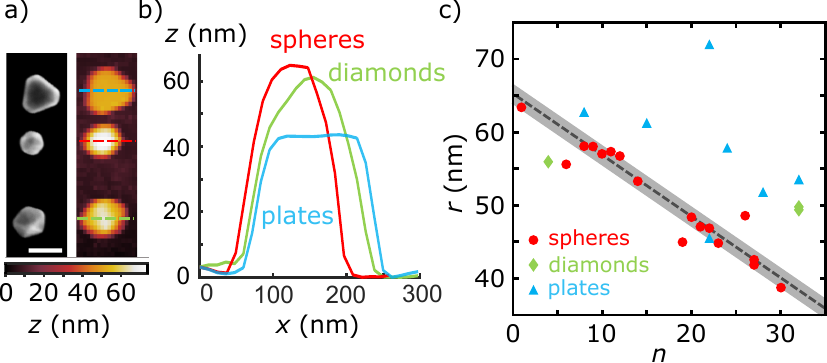}
	\caption{(color online) (a)~SEM (left) and AFM (right) scan of the different particle types present in a mixture with nominal radii of $40$ and $50\,$nm. The white scale bar indicates 100~nm. (b)~AFM scan across particles from sub-figure~(a). (c)~Quantification of the sorting, with colors corresponding to the particle types shown in (b). The spherical particles (red) follow a linear trend of particle radius $r$ and tooth number $n$. The slope of the dashed line was obtained from simulation and its position was shifted in $r$ to obtain a good fit to the data. The shaded area corresponds to a width of $\pm 2\,\sigma_r$. The plates (blue) and diamonds (green) show a different behavior because they experience a reduced interaction energy than spherical particles.}
	\label{fig:fig4}
\end{figure}

After the sorting process, the particles were transported to  compartments R1 to R19 (\fref{fig:fig1}b)) and deposited onto the surface for further inspection, see SM10 for details~\cite{SM}. However, because of the thin polymer film, we observed plastic deformation during the immobilization process. Therefore, we performed a second experiment using a mixture of gold spheres with radii of 80 and 100\,nm  (EM.GC80 and EM.GC100 from BBI solutions, batch  $\# 13063$ and $\# 13083$) and a thick polymer layer. Specifically, the polymer stack consisted of 52\,nm of the adhesion promoter HM8006 (JSR Inc.) and 185\,nm of PPA. The thick polymer layer renders plastic deformation unlikely, as shown recently~\cite{fringes2019Deterministic}.

After sorting, we quantified the particle size using scanning electron microscopy (SEM) and atomic force microscopy (AFM). Similarly as observed previously~\cite{skaug2018nanofluidic}, the particles showed a variety of shapes. As well as spheres, we found flat plates and particles with clearly exposed crystal planes, which we labeled diamonds, see \fref{fig:fig4}a) and b).

\fref{fig:fig4}c) shows the measured effective radius of the particles given by $r^2\pi = A$, where $A$ was determined from SEM images using thresholding, see SM11~\cite{SM} for details. We note, that we treated all systematic errors in SEM imaging by fitting the offset of the line resulting from the model to measured radii of spherical particles. This treatment does not affect our discussion on separation resolution. The spherical particles were well separated with high resolution and follow the predictions of the model (dashed line and shaded area), see SM12~\cite{SM} for details. Counting all spheres, we measure a standard deviation of $2\,$ nm with respect to the model. We note that in our experiments the measured gap distance was stable in time to only 2 nm RMS, partially due to laser noise. This fact will induce hard to predict variations in the travel distance of similar particles.  Given these uncertainties, the experimental data corroborate the simulation results. The plates and diamonds, however, have a flatter shape, and therefore cannot be described by the spherical particle model. The smaller ratio between particle height and diameter reduces their interaction energy with the device surfaces, according to  \eqref{eq:potential}. Consequently, a plate or diamond of the same area as a sphere is transported further into the ratchet.

%
%

%
%
\emph{Conclusion.}---We characterized a nanoparticle sorting device based on a nanofluidic rocking Brownian motor with a linearly increasing tooth height. 
Simulations predict a separation resolution of $\approx 2\,$nm (4 $\sigma$) in radius, which is consistent with the experimentally measured resolution for spherical particles of $2\,$nm (1 $\sigma$) given the experimental conditions. However, particles of different shapes are also transported according to their smallest diameter and therefore cannot be fully separated from the spheres in such a one-dimensional device. Combined with a separation mechanism that differentiates by hydrodynamic radius, a 2D sorting could be implemented that would allow separation by size and smallest particle diameter. 
\eqref{eq:potential} suggests that particles of the same size and different surface potential can also be separated, but with lower resolution because the charge affects only the prefactor and not the exponent. Our modeling shows that, for the $60$-nm particles, a 10\% change in surface potential is required to have the same effect as a 1-nm difference in radius. The method is accordingly separating mainly by size rather than charge.

Similar to conventional devices, the separation resolution is enhanced with increasing external driving force. However, for our devices, not the absolute force but rather the energy per ratchet tooth is important. Thus, higher resolution can be obtained by simply stretching the geometry of the ratchet, and resolutions below 1\,nm would be within reach.

The fast sorting of the particles, achieved in 5--10\,s, is promoted by the small footprint of the device. Therefore, rocking Brownian motors combine high-resolution separation with low applied voltages, high speed, and small device footprints, rendering them ideal for future lab-on-chip applications.
%
%
\newline
\indent
\begin{acknowledgments}
	We thank U.~Drechsler for assistance in fabricating the glass pillars, H.~Wolf for stimulating discussions, 
	and R.~Allenspach and H.~Riel for support. Funding was provided by the European Research Council (StG no.~307079, PoC Grant no.~825794), and the Swiss National Science Foundation (SNSF no.~200021-179148).
\end{acknowledgments}

\bibliographystyle{apsrev4-1}

%
%


\renewcommand{\eqref}[1]{Eq.~(\ref{#1})}
\renewcommand{\fref}[1]{Fig.~\ref{#1}}
\renewcommand{\thefigure}{S\arabic{figure}}
\renewcommand{\theequation}{S\arabic{equation}}

\setcounter{equation}{0}
\setcounter{figure}{0}

\onecolumngrid
\appendix
\newpage

\section{{\large Supplemental Material}}

\section{SM1: Nanoparticles}

We used spherical gold nanoparticles with a nominal diameter of $60\,$nm, $80\,$nm and $100\,$nm obtained from BBI solutions (product codes HD.GC60, HD.GC80, and HD.GC100). The coefficient of variation was~$8$\,\% for all three batches and the particle concentrations were $2.6 \times 10^{10}$ per ml, $1.1 \times 10^{10}$ per ml, and $5.6 \times 10^{9}$ per ml, respectively. For sorting 60\,nm particles into subpopulations, 750$\,\mu$l of the original particle suspension was centrifuged at 7000\,rpm for 10 minutes. The supernatant was removed and replaced by $50\,\mu$l of ultrapure water (Millipore, $18\,$M$\Omega$cm). 
For the experiment on separating $80\,$nm and $100\,$nm gold nanoparticles, we first mixed 
$500\,\mu$l of each of the respective suspensions. Then, we repeated the centrifugation and supernatant replacement step in the same way as for the previous experiment. 
For all experiments, a droplet of $\approx 30$\,$\mu$l of the respective particle suspension was used.

\FloatBarrier
\section{SM2: Experimental Set-Up}

All experiments described in this manuscript were carried out with the nanofluidic confinement apparatus (NCA), which is described in detail in~\cite{Fringes2018}. In short, a nanofluidic gap was formed between a cover slip and the patterned silicon sample. The sample was tilt corrected using three Picomotors (Newport) located under the sample holder until the two confining surfaces were aligned with a precision of 1\,nm across a lateral distance of 10\,$\mu$m. The gap distance itself could be tuned with nanometer-accuracy by moving the cover slip with a linear piezo stage (travel range 100\,$\mu$m, Nano-OP100, Mad City Labs). 
The cover slip comprised a 50\,$\mu$m tall mesa at its center to provide good optical access to the region of interest and an unobstructed approach of the confining surfaces. Around the mesa, four Au electrodes were deposited for the application of electrical fields. For details on the fabrication of the cover slip, please refer to~\cite{Skaug2018}.
Before all experiments, the glass pillar was cleaned with a peel-off polymer  (Red  First Contact, Photonic Cleaning Technologies) and with 30\,s of oxygen and hydrogen plasma at 200\,W (GigaEtch 100-E, PVA TePla GmbH).\\ 
\indent Interferometric scattering detection (iSCAT) was used to image the particles. In detail, we used a green laser (\mbox{532\,nm} continuous wave, Samba 50\,mW, Cobolt) with a beam diameter of $\approx 0.7$\,mm which was focused on the sample by an oil-immersion objective (100x, 1.4 NA, Alpha  Plan-Apochromat,  Zeiss). The focal spot of diameter $\approx 2\,\mu$m was raster-scanned across the field of view  with acousto-optic deflectors (DTSXY, AA Opto-Electronic). Scanning was done with 500\,nm line spacing and 100\,kHz scan rate. Hence, an illuminated nanoparticle interacts with the laser focus of $2\,\mu$m diameter for about 40\,$\mu$s and therefore a particle image constitutes an average over $\approx40\,\mu$s. The reflected light was collected by the same objective and guided onto a high-frame-rate  camera  (MV-D1024-160-CL-12,  Photon  Focus) using a 50:50 beam splitter.

\FloatBarrier
\section{SM3: Sample preparation}

All samples used for our experiments were cut from highly doped silicon wafers. The sample used for sorting 60\,nm gold spheres into subpopulations was first thermally oxidized until an oxide layer of $\approx$ 235\,nm thickness was grown. The thickness of the oxide layer was chosen such that the contrast of the imaged particles became maximal, see \cite{Fringes2016} for details on the underlying optical model. Then, a solution of 10\,vol$\%$ of the adhesion promoter HM8006 (JSR Inc.) was spin coated at 6000\,rpm. Afterwards, the sample was baked for 90\,s at $225^{\circ}\text{C}$ on a hotplate, resulting in a 5\,nm thick highly cross-linked layer of HM8006. In the next step, the sample was coated with a solution of 9 wt\% of the thermally sensitive resist polyphthalaldehyde (PPA) in anisole. The PPA solution was spin coated at 3500\,rpm and cured at $110^{\circ}\text{C}$ for 2\,min, which yielded a 150\,nm thick layer of PPA.\\
\indent
We used thermal scanning probe lithography (t-SPL) to pattern the ratchet structure shown in Fig. 1 of the main text into PPA. The tips were heated to $1000^{\circ}\text{C}$ and capacitively pulled into contact with the PPA by 5 $\mu$s long voltage pulses applied between the tip and the sample. The polymer decomposed locally and evaporated, resulting in a well defined void whose depth can be controlled by the applied voltage. After finishing a line of the pattern, the same tip was used for imaging the written topography. This approach allows for a continuous optimization of the writing parameters (closed-loop lithography) leading to a vertical patterning accuracy of 1\,nm~\cite{Pires2010}.\\
\indent 
In the next step, the t-SPL written pattern was transferred into the oxide layer by reactive ion etching, using a mixture of the gases CHF$_3$, Ar and O$_2$. Given the etch-selectivity between polymer and SiO$_2$ of 2:1, the maximum depth of the patterns in SiO$_2$ reduced from 120\,nm to 60\,nm. Finally, the sample was spin coated with OTL at $3500$\,rpm and baked for 120\,s at $240^{\circ}\text{C}$ on a hotplate resulting in a $10$\,nm thick layer of OTL.\\
\indent 
The sample used for separating 80\,nm and 100\,nm gold nanoparticles was fabricated in a slightly different way. First, it was spin coated with the undiluted, pre-formulated solution of HM8006 at 6000\,rpm. Then, the sample was baked at $225^{\circ}\text{C}$ on a hotplate for 90\,s for cross-linking, yielding a $\approx 52$\,nm thick layer of HM8006 (JSR Inc.). Next, the sample was spin coated with a pre-formulated solution of PPA (AR-P 8100.06, ALLRESIST GmbH) at 1750\,rpm. It was cured at $110^{\circ}\text{C}$ for 2\,min to evaporate residual solvent which resulted in a $\approx$ 185\,nm thick layer of PPA. In the last step, the sorting ratchet was patterned into the PPA using t-SPL as already described above.

\FloatBarrier
\subsection{SM4: Stability of the experimental apparatus}

\begin{figure}[ht]
	\centering
	\includegraphics[width=0.8\textwidth]{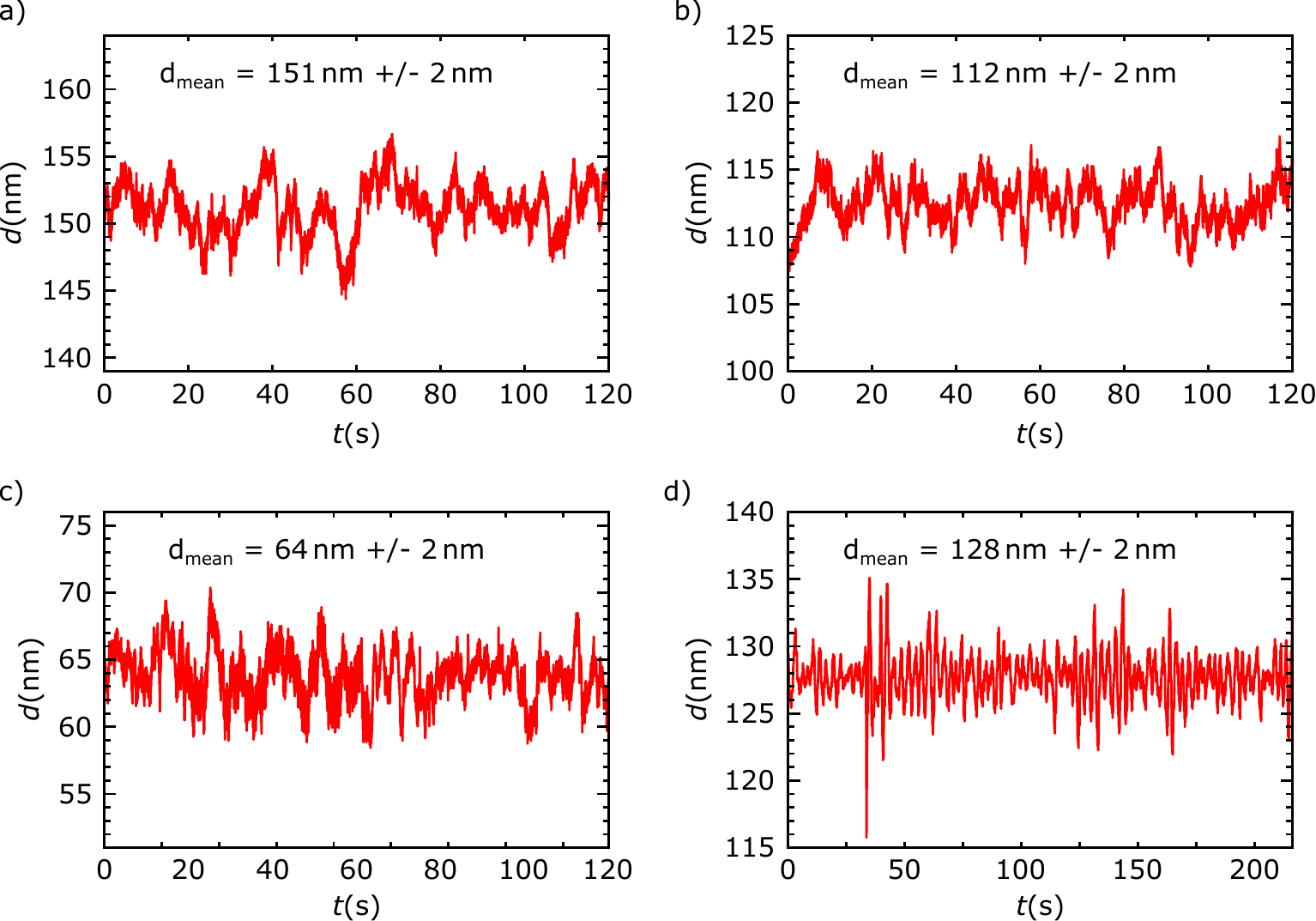}
	\caption{Particle sorting with nm resolution requires a high stability of the nanofluidic gap.
		a) and b) Gap distance fluctuations when measuring the potential landscape 
		experienced by 60\,nm particles. c) Gap stability during sorting 60\,nm gold nanoparticles into subpopulations. d) Corresponding plot for separating 80\,nm and 100\,nm particles.
		As can be seen, the standard deviation for all experiments was 2\,nm corresponding to a height difference of approximately two neighboring teeth of the sorting ratchets. Note that the y-axis in all sub-figures encompasses 25\,nm.}
	\label{fig:fig2_som}
\end{figure}

\FloatBarrier
\section{SM5: Measuring diffusion, force and interaction potential}

\begin{figure}[ht]
	\centering
	\includegraphics[width=0.8\textwidth]{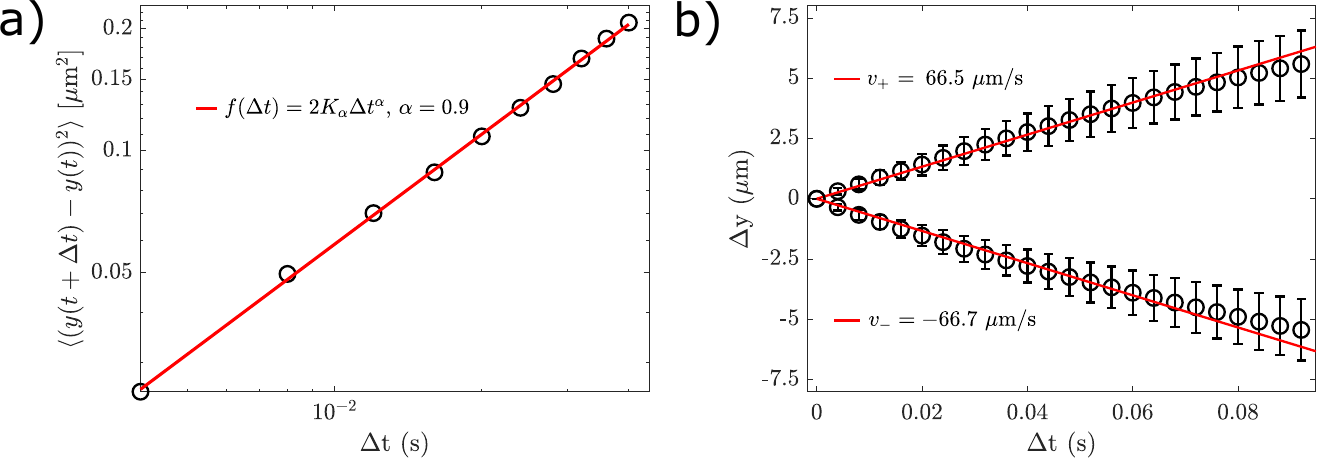}
	\caption{a) The diffusion constant was inferred from the mean square displacements of particles along the y-axis of the drift field D1. The experimental data was fitted with a power law function. The exponent $\alpha = 0.9$ indicates subdiffusion, while the prefactor corresponds to $2K_\alpha$, where $K_\alpha$ is a generalized diffusion coefficient. b) Determination of the particle drift velocity in the drift field for an applied AC voltage of 3.5\,V at 5\,Hz. The displacements are averages over all particles and trajectories.}
	\label{fig:fig3_som}
\end{figure}

\noindent\textbf{Determination of the diffusion coefficient:} Before the AC electric fields were switched on to start the sorting process, the particles were allowed to diffuse freely for $\approx 20$\,s. By tracking the particles that were trapped in the drift field D1 (see Fig. 1b) of the main text) we could infer the free diffusion coefficient. Therefore, we first calculated the mean-squared displacement along the y-axis of the drift field
\begin{equation}
\langle \Delta y^2(\Delta t)\rangle = \left\langle \frac{1}{N-1}\sum\limits_{i=1}^{N-1}(y(t_i + \Delta t)-y(t_i))^2 \right\rangle
\end{equation}
for different time intervals $\Delta t$ where $N$ is the number of observed positions per trajectory and $\langle ... \rangle$ represents the ensemble average. As shown in \fref{fig:fig3_som} a), the inferred values can be fitted by a power-law function of the form $\langle y^2(\Delta t)\rangle = 2K_{\alpha}\Delta t^{\alpha}$ where $K_{\alpha}$ is a generalized diffusion coefficient and ${\alpha}$ the anomalous diffusion exponent. At the gap distances used during our experiments, we have $0 < {\alpha} < 1$ which is called the subdiffusive regime~\cite{Fringes2018}.
In this regime, the diffusion coefficient becomes time-dependent and is best described by 
\begin{equation}
D_{\alpha}(\Delta t) = \frac{\langle \Delta y^2(\delta t)\rangle}{2 \Delta t} = K_{\alpha} \Delta t^{\alpha-1}.
\end{equation}
Then, for further calculations, we used the value of the diffusion coefficient at the shortest experimental timescale $D_0\approx D_{\alpha}(\Delta t_{\rm min}) = K_{\alpha}\Delta t_\text{min}^{{\alpha}-1} = K_{\alpha}\left(\frac{1}{\text{fps}}\right)^{{\alpha}-1}$, where $1/\Delta t_{\rm min} ={\rm fps}$ is the frame rate of the camera. For the sorting experiment of 60\,nm particles shown in Fig. 2 of the main text, we measured a diffusion coefficient ${D_0 = 3.2 \pm 0.2\,\mu{\rm m}^2/{\rm s}}$.\\ 

\noindent\textbf{Force measurement:} In order to infer the rocking force $F(t)$ experienced by the gold nanoparticles, it was necessary to measure their drift speed when the AC field along the y-axis was switched on. Therefore, the motion of the particles trapped in the drift field D1 was tracked during the sorting process. The recorded trajectories were split each time the applied voltage changed its sign. Then the average displacement along the y-axis was calculated for both orientations of the electric field, see ~\fref{fig:fig3_som} b). The average drift velocity in forward and backward direction was then determined by fitting a linear curve to the displacement values. The fit to the experimental datapoints was weighted with the inverse of the standard deviation obtained across the different trajectories. Finally, the applied rocking force could be inferred by combining Einstein's relation $D_0 = k_{\rm B}T/\gamma$ and Stokes' equation $F_\text{drag} = \gamma \langle v_\text{drift}\rangle$ to 
\begin{equation}
F = k_{\rm B}T \langle v_\text{drift}\rangle/D_0.
\end{equation}

\noindent Using the measured drift velocity and diffusion coefficient, the applied force for the experiment shown in Fig. 2a) of the main text was $20.7 \pm 1.3 \,k_{\rm B}T$.

\noindent\textbf{Experimental Potential Landscapes:} \noindent In a nanofluidic gap, electrostatic interactions between the charged surfaces of the nanoparticles and the confining walls create a potential landscape that varies as a function of local gap distance. 

\begin{figure}[ht]
	\centering
	\includegraphics[width=0.25\textwidth]{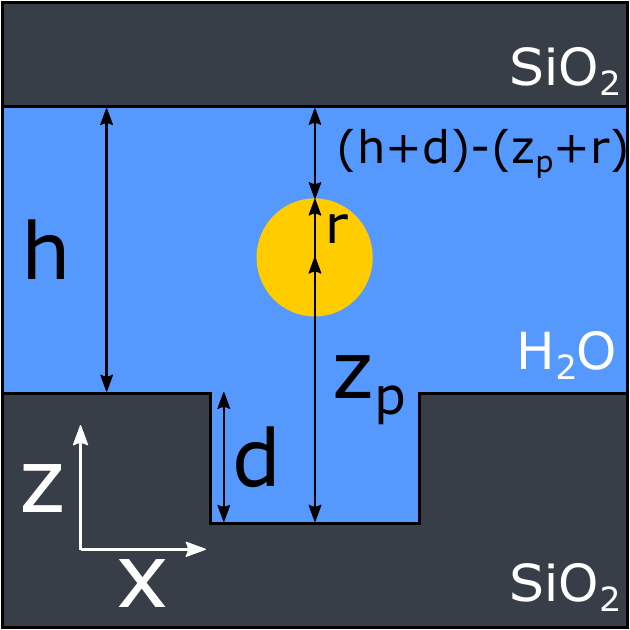}
	\caption{Schematic of a gold nanoparticle in a patterned nanofluidic gap.}
	\label{fig:schematic}
\end{figure}

\noindent Let us consider a nanofluidic gap of height $h$ with one of the confining walls patterned with a surface topography of depth $d(x,y)$ and a spherical nanoparticle of radius $r$ situated at a vertical distance $z_p$ from the lower confining surface. For our surface patterns with sidewall slopes of $<1$, the electrostatic potential can be approximated using the sphere-plane interaction model \cite{Skaug2018, Behrens} as
\begin{equation}~\label{electrostatic_potential}
U(x,y) = \psi_{\text{p,s,eff}}\,r\left(\text{exp}(-\kappa(z_p-r)) + \text{exp}(-\kappa(h+d(x,y)-z_p - r))\right)
\end{equation}
\noindent with $\psi_{\text{p,s,eff}}$ representing an effective surface potential of the linearized Poisson-Boltzmann equation for the nanoparticle and the confining surfaces. The schematic in \fref{fig:schematic} illustrates this situation. 


\noindent The Boltzmann relation states that the probability of finding a particle at a location ($x,y,z$) depends exponentially on the potential at this location $U(x,y,z)$:

\begin{equation}
p(x,y,z) \propto \text{exp}(-U(x,y,z)/kT).
\end{equation} 

To experimentally determine the potential landscape in our device, we trapped nanoparticles in the ratchet at a gap distance at which the entire structure was probed. Then, we let the particles diffuse for 120\,s and recorded images at 250\,fps. We obtained a 1-D probability distribution $P(x)$ by averaging the tracked particle positions along the y-axis and calculating the frequency of observing a particle at position $x$. These probabilities were then converted to potential differences:

\begin{equation}~\label{eq:prob}
\frac{\Delta U(x)}{kT} = -\text{ln}\left(\frac{P(x)}{P_{\text{max}}(x)}\right).
\end{equation}\\

\noindent The NCA allowed us to repeat this measurement at different gap distances with the same ratchet structure and the same particles. We probed the ratchet before the sorting experiment shown in Fig.~2a) of the main text at three gap distances: 151\,nm, 112\,nm and 103\, nm. These measurements were used to extract the effective surface potential $\psi_{\text{p,s,eff}}$ and the Debye length $\kappa^{-1}$, by finding the best fit of calculated potentials to the experimental data at the different gap distances, see discussion below.\\

\begin{figure}[ht]
	\centering
	\includegraphics[width=0.65\textwidth]{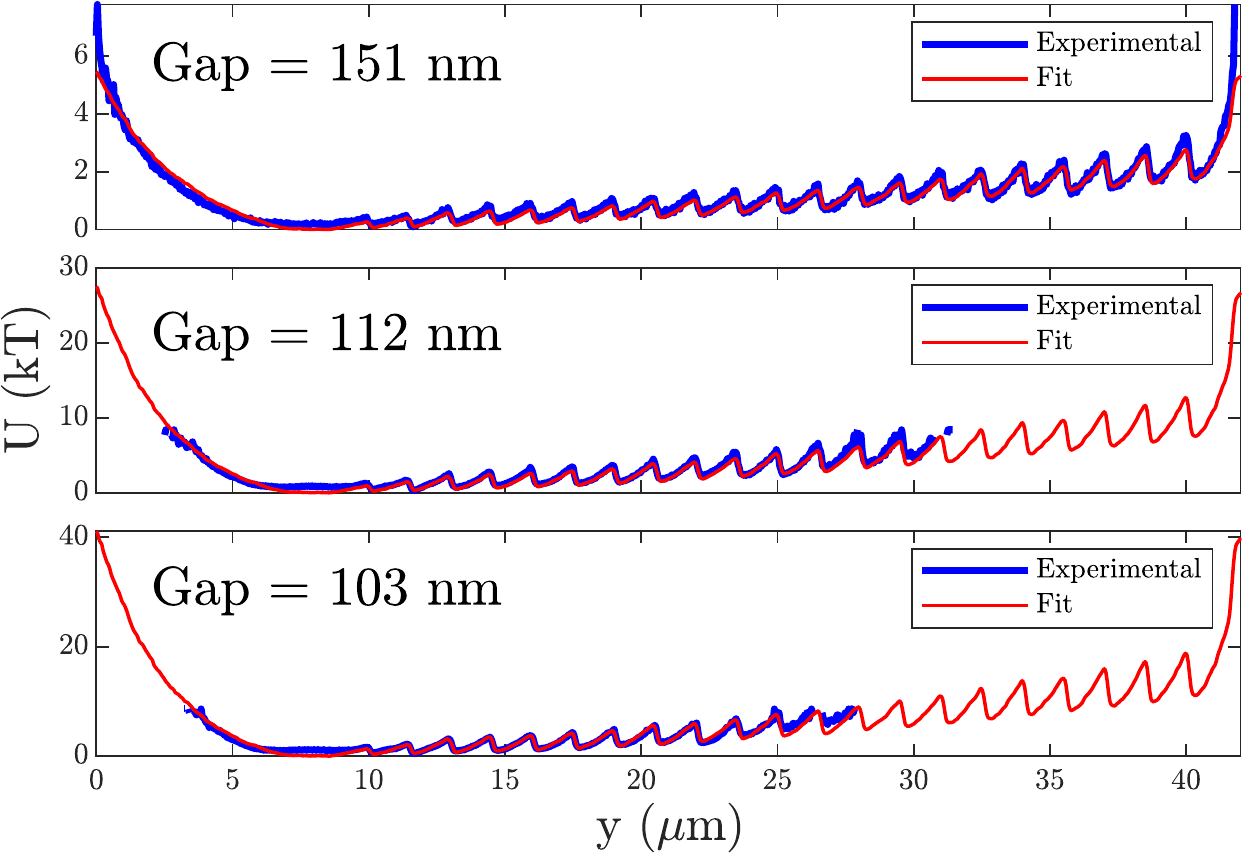}
	\caption{Experimental potentials measured at three gap distances. A Levenberg-Marquardt nonlinear regression was used to globally fit the two parameters $\kappa$ and $\psi_{\text{p,s,eff}}$ to equation~\ref{eq:integration}.}
	\label{fig:fig4_som}
\end{figure}

\noindent We used the topography $d(x)$ of the ratchet, as measured by AFM, to calculate a 2D probability distribution $P(x,z)$ using \eqref{eq:prob}. To derive the corresponding 1D probability distribution along the x-axis of the ratchet, an integration along the z-axis of the nanofluidic gap had to be performed:
\begin{equation}~\label{eq:integration}
P(x) = \int_{r}^{d(\text{x})-r}p(x,z)\text{d}z = \int_{r}^{d(\text{x})-r}C\text{exp}\left(-\dfrac{\psi_{\text{p,s,eff}}\,r\left(\text{exp}(-\kappa(z-r)) + \text{exp}(-\kappa(h+d(x)-z - r))\right)}{kT}\right)\text{d}z 
\end{equation}
where $C$ is a normalization factor. $P(x)$ was then converted to a potential landscape using equation~\ref{eq:prob}. A fit of this function to the experimental data was performed with a Levenberg-Marquardt nonlinear regression, using a nominal particle radius of 30\,nm with fitting paramters $\psi_{\text{p,s,eff}}$ and $\kappa$. We performed a global fit over the three measurements carried out at gap distances between 103\,nm and 151\,nm, as shown in \fref{fig:fig4_som}. We found a unique pair of values for $\kappa$ and $\psi_{\text{p,s,eff}}$ that fitted the measured data best.

From the experimental data, we only used the particle locations inside the ratchet region, additionally neglecting the two first and the two last ratchet teeth. This avoided errors from erroneous particle tracking due to crowding (two first teeth) or errors due to sparse statistics (two last teeth). From this global fit, we obtained values of $\kappa^{-1} = 10.8\,\pm\,0.1\,$nm and $\psi_{\text{p,s,eff}} = 5.3\,\pm\,0.2\,k_{\rm B}T$.\\

\FloatBarrier
\section{SM6: Numerical solution of the Fokker-Planck equation}

The dynamics of a ratchet with potential $V(x)$ and time dependent external rocking force $F(t)$ can be described in terms of a probability density $\rho(x,t)$ which obeys the Fokker-Planck equation
\begin{equation}
\partial_t \rho(x,t) = \partial_x \left[ \left( \frac{1}{\gamma}\partial_x \tilde{V}(x,t) + D_0 \partial_x \right) \rho(x,t) \right]
\label{eq:fokkerplanck}
\end{equation}
with drag constant $\gamma$, diffusion coefficient $D_0$ and $\tilde{V}(x,t) = V(x) + x F(t)$.
Mathematically, the sorting process constitutes an initial value problem where the probability density $\rho_0(x)$ at $t=0$ evolves into a new probability density $\rho(x)$ for ${t>0}$.
In general, such an initial value problem can only be solved numerically. Therefore, we discretized \eqref{eq:fokkerplanck} with respect to $x$ and approximated the first and second spatial derivatives by 
finite differences. As both the potential $V(x)$ and the probability density $\rho(x,t)$ exhibit steep
gradients, it was necessary to use finite differences with high accuracy to guarantee both numerical
stability and a sufficient precision of the numerical solution. Therefore, we used the central finite differences
\begin{eqnarray}
\partial_x f(x) &\approx&  \frac{-f(x_{i-3})+9f(x_{i-2})-45f(x_{i-1})+45f(x_{i+1})-9f(x_{i+2})+f(x_{i+3})}{60 \Delta x} + O(\Delta x^6) \label{eq:dfdx}\\
\partial_x^2 f(x) &\approx& \frac{2f(x_{i-3})-27f(x_{i-2})+270f(x_{i-1})-490f(x_{i})+270f(x_{i+1})-27f(x_{i+2})+2f(x_{i+3})}{180\Delta x^2} + O(\Delta x^6) \label{eq:df2dx2}
\end{eqnarray}
where $x_i$ is the $i$th supporting point ($i \in [1,N]$) and $\Delta x$ the spacing between neighboring supporting points. At the boundaries of the ratchet we resorted to corresponding forward and backward finite difference formulas to calculate the spatial derivatives. As a result, \eqref{eq:fokkerplanck} was transformed into a system of $N$ ordinary differential equations which could then be solved using standard numerical routines like e.g. scipy.integrate.solve\_ivp from the scipy package for Python.
\begin{figure}[ht]
	\centering
	\includegraphics[width=0.8\textwidth]{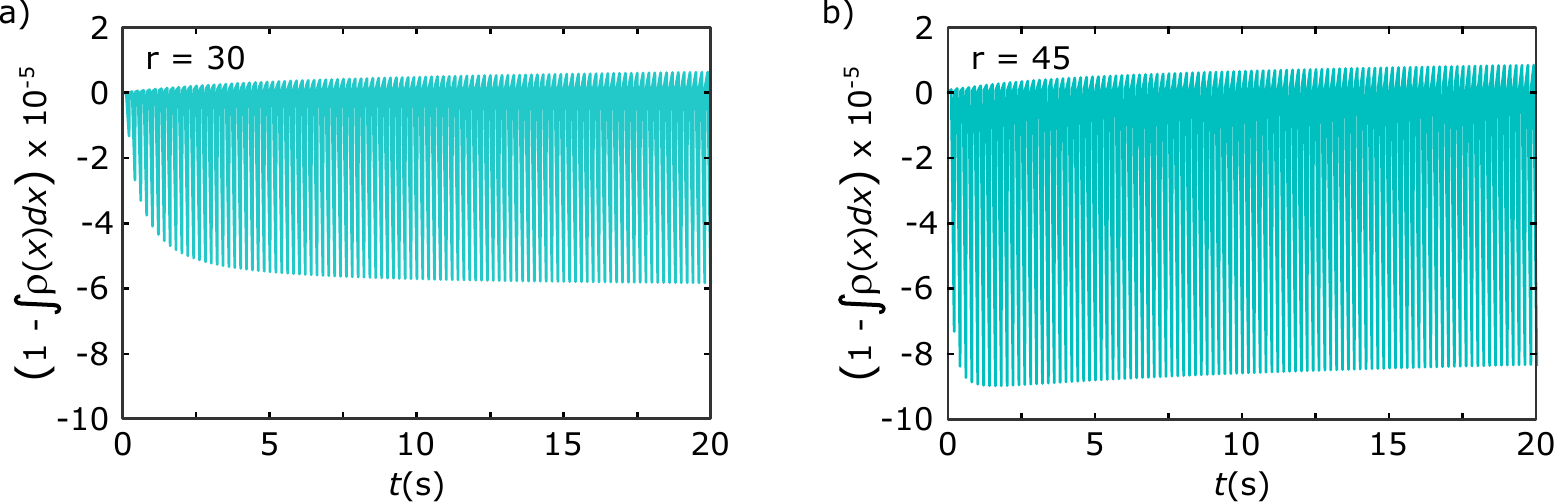}
	\caption{a) Conservation of the probability density when simulating the transport of a particle with radius $r=30$\,nm inside the sorting ratchet shown in Fig. 1a) of the main text. b) Corresponding plot for a particle with radius $r=45$\,nm that is transported by the ratchet shown in \fref{fig:80nm_som}a). As can be seen, the probability density is conserved up to a numerical error of~$\leq 10^{-4}$.}
	\label{fig:fig1_som}
\end{figure}
\newline
\indent
As the number of particles must stay constant during the sorting process, the probability current 
\begin{eqnarray}
S(x,t)  =  -\left( \frac{1}{\gamma} \partial_x \tilde{V}(x,t) + D_0 \partial_x \right) \rho(x,t)
\end{eqnarray}
has to vanish at the boundaries. We enforced these boundary conditions by setting $\rho(x_0)$ and $\rho(x_N)$ at every iteration step such that the conditions $S(x_1,t) = 0$ and $S(x_N,t) = 0$ were fulfilled.
\newline
\indent
We observed that the quality of the numerical solution of the initial value problem depends critically on choosing a sufficiently fine grid. In our simulations we used a spatial resolution of $\Delta x\leq 6\,$nm which is much smaller than the particle size and turned out to deliver results with good accuracy as shown in \fref{fig:fig1_som}.

\FloatBarrier
\section{SM7: Details on simulating the sorting process of 60\,nm gold nanoparticles}

\begin{figure}[ht]
	\centering
	\includegraphics[width=0.8\textwidth]{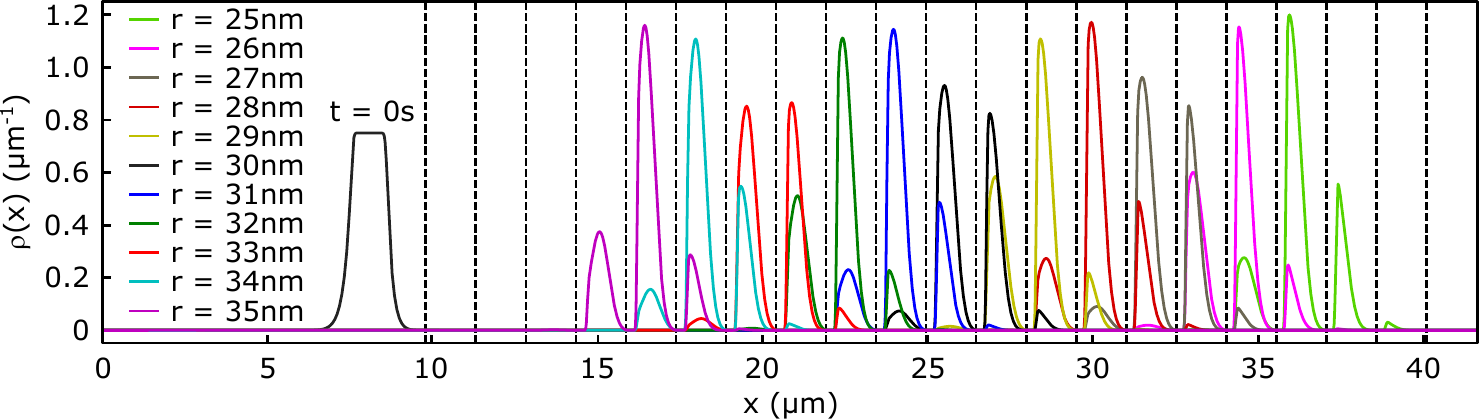}
	\caption{Simulated probability densities for gold nanoparticles of different size after sorting. For particles with a radius of 30\,nm, we used the potential shown in Fig. 1d) of the main text, a rocking force of $20.7$\,kT/$\mu$m, and a diffusion coefficient of $D_0 = 3.22 \mu {\rm m}^2/{\rm s}$. For particles with different radius, we rescaled the respective quantities correspondingly.}
	\label{fig:fig10_som}
\end{figure}

According to BBI solutions, the average particle radius of batch $\# 19080123$ of EM.GC60  was $30.25$\,nm with a coefficient of variation of~$8$\,\%. To assess the particle sorting performance of the device shown in Fig. 1c) of the main text, we assumed a Gaussian distribution of integer-valued particle radii and considered only populations which contributed at least $1.5\,\%$ to the overall distribution, see \tref{tab:tab1_som}.
\begin{center}
	\begin{table}[ht]
		\begin{tabular}{  l | c | c | c | c | c | c | c | c | c | c | c | c | c | c | c }
			\hline
			radius $R$ & 24\,nm & 25\,nm & 26\,nm & 27\,nm & 28\,nm & 29\,nm & 30\,nm & 31\,nm & 32\,nm & 33\,nm & 34\,nm & 35\,nm & 36\,nm & 37\,nm & $\sum$ \\ \hline\hline
			
			fraction & 0.6\,$\%$ & 1.6\,$\%$ & 3.6\,$\%$ & 6.7\,$\%$ & 10.7\,$\%$ & 14.3\,$\%$ & 16.3\,$\%$ & 15.6\,$\%$ & 12.6\,$\%$ & 8.7\,$\%$ & 5.0\,$\%$ & 2.4\,$\%$ & 1.0\,$\%$ & 0.4\,$\%$ & 99.5\,$\%$ \\ 
			
			simulation & 0\,$\%$ & 1.6\,$\%$ & 3.7\,$\%$ & 6.9\,$\%$ & 10.9\,$\%$ & 14.7\,$\%$ & 16.7\,$\%$ & 16.0\,$\%$ & 13.0\,$\%$ & 8.9\,$\%$ & 5.1\,$\%$ & 2.5\,$\%$ & 0\,$\%$ & 0\,$\%$ & 100\,$\%$\\ 		
			\hline		
		\end{tabular}	
		\caption{Fraction of particles with radius $R \pm 0.5$\,nm of the overall population (upper row) and particle radii used for simulating the sorting process (bottom row). Note, that the respective fractions in the bottom row were re-normalized for our simulation such that they summed up to $100\,\%$ again.}
		\label{tab:tab1_som}
	\end{table}
\end{center}
Then, we simulated the time propagation of particles with a radius between 25\,nm and 35\,nm, see \fref{fig:fig10_som}, and weighted the obtained probability densities according to \tref{tab:tab1_som}. Using this approach, we could easily calculate the probability distributions shown in Fig. 2a) and 2c) of the main part.

\FloatBarrier
\newpage
\section{SM8: Drift speed in the sorting device for separating 60\,nm particles}

\begin{figure}[ht]
	\centering
	\includegraphics[width=0.4\textwidth]{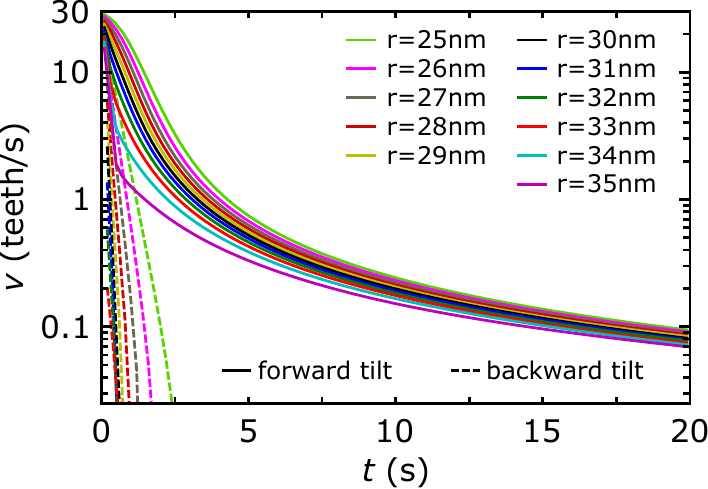}
	\caption{The resolution of the sorting device is mainly governed by the forward current ($F < 0$) as the backward current ($F > 0$) is negligible for all particles after already 2.5\,s of sorting. }
	\label{fig:drift_speed_som}
\end{figure}

\FloatBarrier
\section{SM9: Details on the calculation of the sorting resolution}

\begin{figure}[ht]
	\centering
	\includegraphics[width=0.8\textwidth]{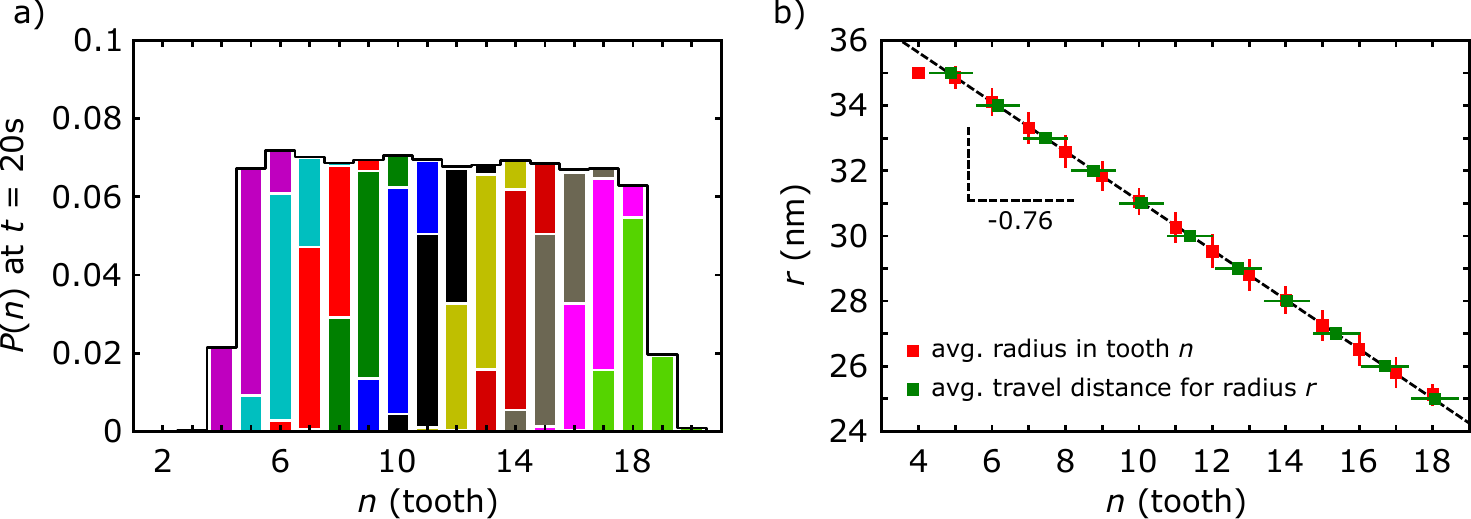}
	\caption{a) Simulated propability distribution of gold nanoparticles after 20\,s of sorting assuming equal weights for all particle populations. Note, that we used the same color code as in \fref{fig:fig10_som}. b) Expected tooth numbers $\langle n \rangle$ for particles of radius $r$ (green squares) and average radius $\langle r\rangle$ for particles found in tooth $n$ (red squares).}
	\label{fig:figSR_som}
\end{figure}

In order to determine the achievable sorting resolution, we again considered the time propagation as already discussed in section SM7. However, here, we assigned a constant weight of $9.1\,\%$ to all 11 particle populations with radii between $25\,$nm and $35\,$nm. Then, we calculated the expected particle distribution in the sorting ratchet after 20\,s of rocking, see \fref{fig:figSR_som}a). One can clearly observe that particles of the same size spread over several teeth of the ratchet and that in each tooth there is a mixture of particles with different radii. Hence, from the result shown in \fref{fig:figSR_som}b), we could calculate the average radius $\langle r \rangle$ of a particle found in ratchet tooth $n$ at the end of the sorting process. Similarly, we could calculate in which tooth a particle of radius $r$ is expected to be found, see \fref{fig:figSR_som}b). Clearly, both the expected tooth number $\langle n \rangle$ and the average radius $\langle r \rangle$ exhibit the same linear behavior. Note, that the slope $m=-0.76$\,nm$/$tooth of the dashed line agrees with the ratio of the standard deviations $\Delta r = 0.48$\,nm and $\Delta n = 0.64$\,tooth. In Fig. 3b) and 3d) of the main text, we used the slope $m$ to convert the standard deviation expressed in teeth into an nm value.

\FloatBarrier
\section{SM10: Final state of sorting 60\,nm gold nanoparticles}

\begin{figure}[ht]
	\centering
	\includegraphics[width=0.8\textwidth]{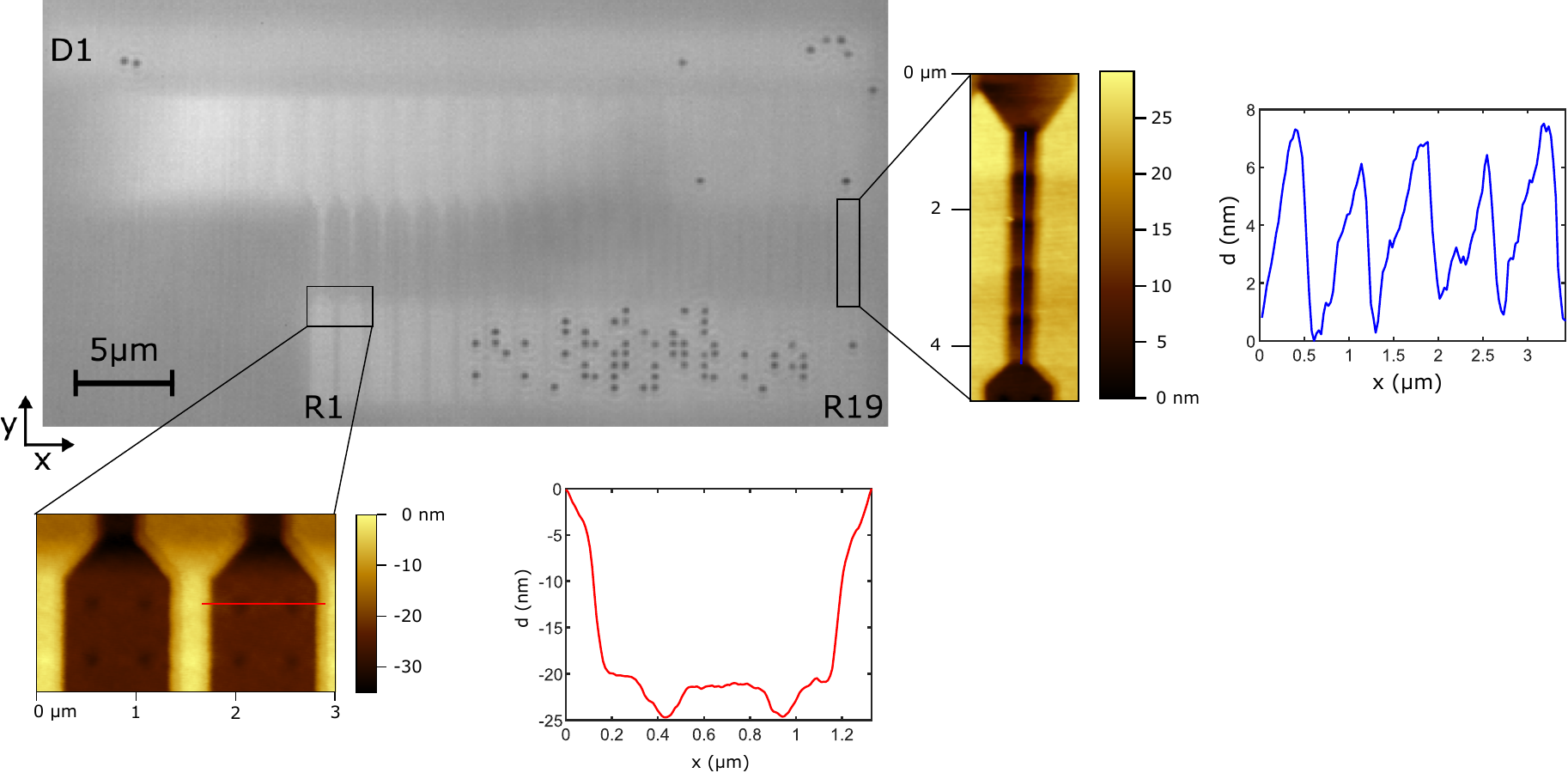}
	\caption{At the end of the sorting process, the particles were transported to the reservoirs R1 to R19 by switching on an AC voltage of 5.5\,V at 10 Hz along the y-axis. Inside the reservoirs, the particles were immobilized in traps that were arranged along a regular grid.}
	\label{fig:lastframe}
\end{figure}

After $\approx 20$\,s of sorting, the particle distribution was practically stable and we switched off the AC voltage along the x-axis. Then, a 5.5\,V AC voltage at 10\,Hz was applied along the y-axis. The particles were transported to their respective reservoirs (R1-R19) through 500\,nm narrow channels by intercalated ratchets, see \fref{fig:lastframe}. Each reservoir contained $\approx 5$\,nm deep electrostatic traps arranged along a regular grid. The traps could only be occupied by single particles
which allowed us to keep particles apart from each other for subsequent measurements, see \fref{fig:lastframe} for an AFM scan of the topography of the trap.

\FloatBarrier
\newpage
\section{SM11: Determination of particle size}

\begin{figure}[ht]
	\centering
	\includegraphics[width=0.8\textwidth]{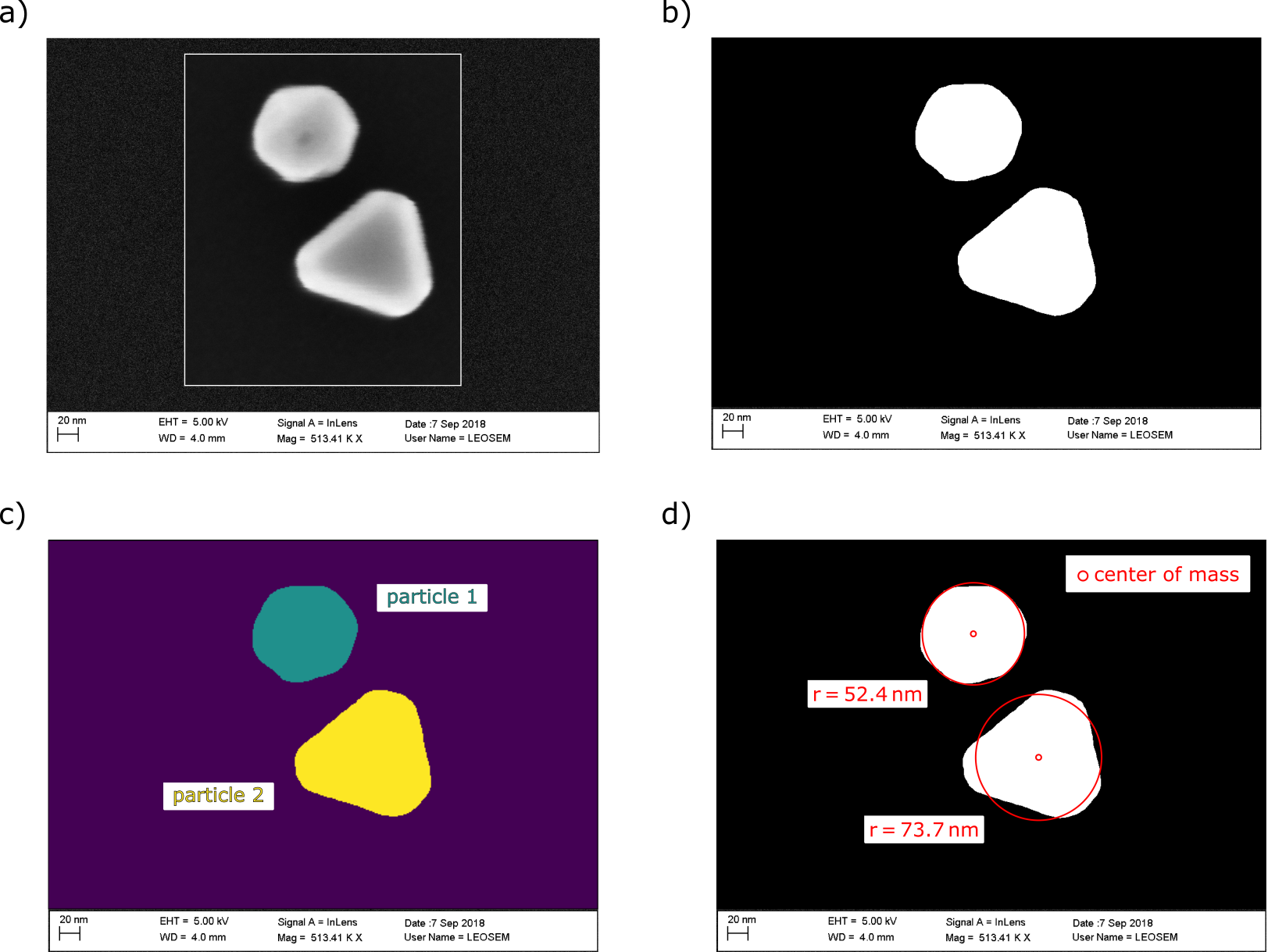}
	\caption{Determination of the effective particle radius $r_{\rm eff}$ from an SEM image. a) Raw image recorded with a Leo 1550 SEM. b) Rescalded image after filtering and thresholding. c) Detection of connected components. d) Image b) together with the associated effective particle size for both particles.}
	\label{fig:fig20_som}
\end{figure}

To determine the size of the deposited particles we used a scanning electron microscope (SEM) of type
Leo 1550. In detail, we applied the in-lens detector of the SEM at a magnification of $\approx 500000$ to 
record pictures of the particles, see \fref{fig:fig20_som}a). 
Then, we normalized the contrast of the
recorded image to 1 and applied a Gaussian filter with a width of one pixel to reduce random detector noise. All pixels with a relative contrast $>0.4$ were attributed to a particle whereas pixels
with a relative contrast $\leq 0.4$ were attributed to the background, see \fref{fig:fig20_som}b).
In the next step, we determined connected components using the function scipy.ndimage.label from the
scipy package for Python and associated each connected component with a particle, see \fref{fig:fig20_som}c).
As the pixel size was stored together with the image, the visible area $A_{\rm vis}$ of each particle could be
determined by counting the number of pixels belonging to the particle. In the last step, an effective
radius $r_{\rm eff}$ was attributed to the particle according to $A_{\rm vis} = \pi r_{\rm eff}^2$, see \fref{fig:fig20_som}d).

\FloatBarrier
\newpage
\section{SM12: Details on separating 80\,nm and 100\,nm particles}

In this experiment the parameters were slighly different from the experiment described in the main text. The total height differnce in the ratchet was 34\,nm with a difference of 1.9\,nm between teeth. The Debye length and the diffusion coefficients were adapted from previously measured parameters on the same sample stack \cite{Skaug2018}: The Debye length was 18\,nm and the diffusion coefficient was 4.6$\,\mu$m$^2$/s. The gap distance determined at the deepest position of the ratchet was 205\, nm, the force for a 30\,nm diameter sphere was 29.25 $k_B\,T/\mu$m and the rocking frequency 5\,Hz.
The graphs shown below depict the results corresponding to the figures in the main text for the aforementioned parameters. 

\begin{figure}[ht]
	\centering
	\includegraphics[width=0.8\textwidth]{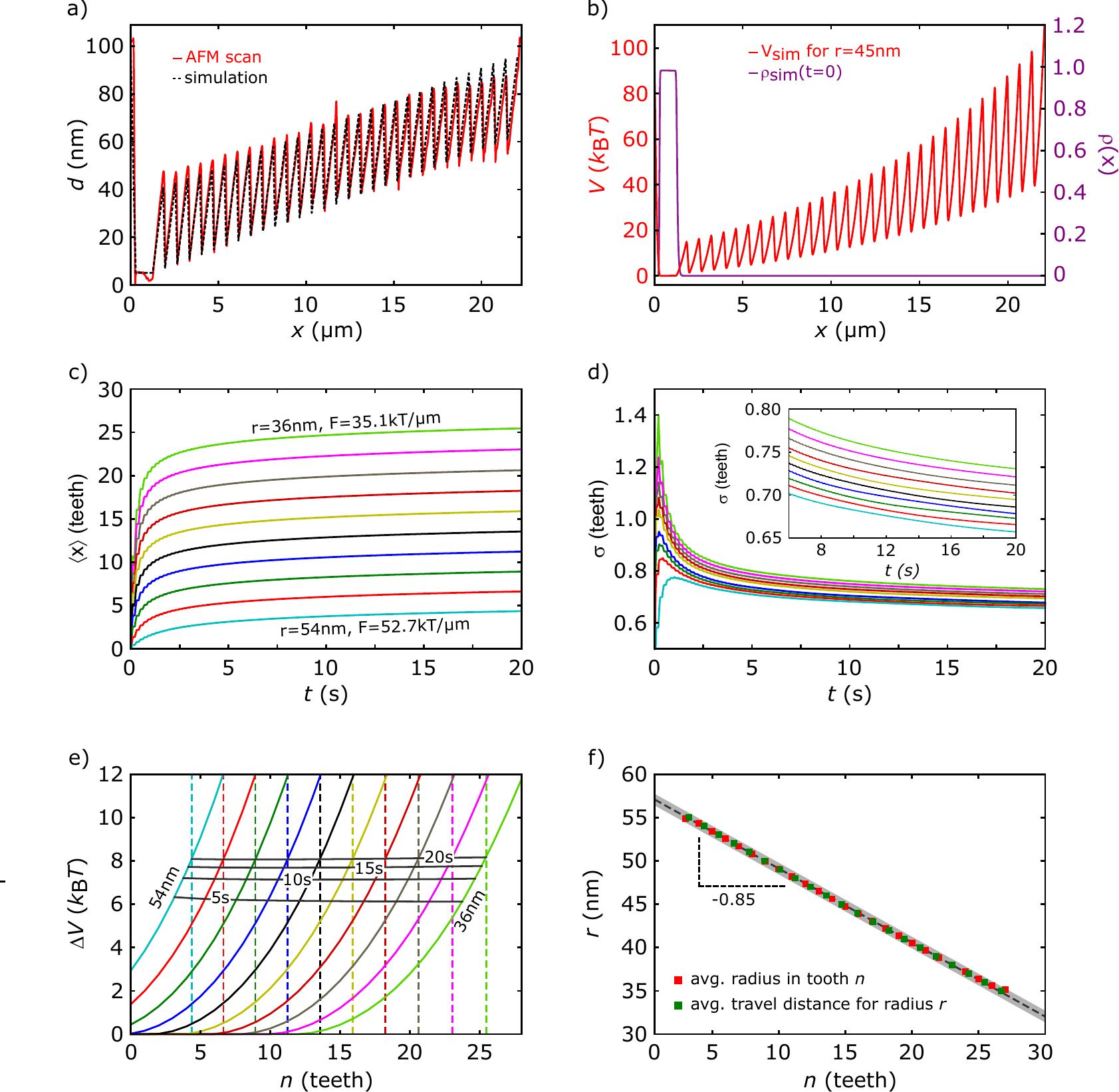}
	\caption{a) Comparison between the topography of the sorting ratchet used for separating 80\,nm and 100\,nm as obtained from an AFM scan (solid red line) and the topography used for our simulation (dashed black line). b) Interaction potential experienced by a particle of radius $r=45$\,nm for the simulated topography shown in a) together with the initial probability density. c) Expected average travel distance for particles of radii between $36\,$nm and $54\,$nm (in steps of 2\,nm) as a function of time and for rocking forces between $35.1\, k_B T / \mu {\rm m}$ and $52.7\, k_B T / \mu {\rm m}$. Note that the same color code is used through sub-figures c) - e). d) Standard deviation for the curves shown in sub-figure c) as a function of time. After $\approx 5$\,s the standard deviation reaches a practically constant value of $\approx 0.75$\,teeth. e) Remaining potential barriers after 5\,s, 10\,s, 15\,s, and 20\,s of sorting. f) Expected tooth numbers $\langle n \rangle$ for particles of radius $r$ (green squares) and average radius $\langle r\rangle$ for particles found in tooth $n$ (red squares). Note that we assumed a constant weight of $4.78\,\%$ for all 21 considered particle populations with radii between $35\,$nm and $55\,$nm. The gray shaded area corresponds to an error of $1\,\sigma$.}
	\label{fig:80nm_som}
\end{figure}

\FloatBarrier

\end{document}